\documentclass[fleqn,10pt]{wlscirep}
\usepackage[utf8]{inputenc}
\usepackage[T1]{fontenc}
\usepackage{lineno}
\usepackage[utf8]{inputenc}
\usepackage[T1]{fontenc}
\usepackage{lineno}
\usepackage{float}
\usepackage{amsmath,amssymb,amsfonts}
\usepackage{array,multirow,graphicx}
\usepackage{float}
\usepackage{algorithmic}
\usepackage{graphicx}
\usepackage{textcomp}
\usepackage{xcolor}
\usepackage{hyperref}
\usepackage{cleveref}
\usepackage{footmisc}

\usepackage{xcolor,colortbl}

\definecolor{Gray}{gray}{0.85}
\definecolor{LightCyan}{rgb}{0.88,1,1}

\newcolumntype{a}{>{\columncolor{Gray}}c}
\newcolumntype{b}{>{\columncolor{white}}c}

\title{VinDr-CXR: An open dataset of chest X-rays with radiologist's annotations}

\author[1,2,$\dag$]{Ha Q. Nguyen}
\author[4,$\dag$]{Khanh Lam}
\author[5,$\dag$]{Linh T. Le}
\author[1,2,3,*]{Hieu H. Pham}
\author[1]{Dat Q. Tran}
\author[1]{Dung B. Nguyen}
\author[4,$\ddag$]{Dung D. Le}
\author[4,$\ddag$]{Chi M. Pham}
\author[4,$\ddag$]{Hang T. T. Tong}
\author[4,$\ddag$]{Diep H. Dinh}
\author[4,$\ddag$]{Cuong D. Do}
\author[5,$\ddag$]{Luu T. Doan}
\author[5,$\ddag$]{Cuong N. Nguyen}
\author[5,$\ddag$]{Binh T. Nguyen}
\author[5,$\ddag$]{Que V. Nguyen}
\author[5,$\ddag$]{Au D. Hoang}
\author[5,$\ddag$]{Hien N. Phan}
\author[5,$\ddag$]{Anh T. Nguyen}
\author[6,$\ddag$]{Phuong H. Ho}
\author[1]{Dat T. Ngo}
\author[1]{Nghia T. Nguyen}
\author[1]{Nhan T. Nguyen}
\author[1]{Minh Dao}
\author[1,7]{Van Vu}
\affil[1]{Smart Health Center, VinBigData JSC,  Hanoi, Vietnam}
\affil[2]{College of Engineering and Computer Science, VinUniversity,  Hanoi, Vietnam}
\affil[3]{VinUni-Illinois Smart Health Center, VinUniversity,  Hanoi, Vietnam}
\affil[4]{Hospital 108, Department of Radiology, Hanoi, Vietnam}
\affil[5]{Hanoi Medical University Hospital, Department of Radiology, Hanoi, Vietnam}
\affil[6]{Tam Anh General Hospital, Department of Radiology, Ho Chi Minh City, Vietnam}
\affil[7]{Yale University, Department of Mathematics, New Heaven, CT 06511, U.S.A.}
\affil[*]{corresponding author: Hieu H. Pham (hieu.ph@vinuni.edu.vn)}
\affil[$\dag$]{these authors contributed equally to this work}
\affil[$\ddag$]{these authors contributed equally to this work}

\begin{abstract}
Most of the existing chest X-ray datasets include labels from a list of findings without specifying their locations on the radiographs. This limits the development of machine learning algorithms for the detection and localization of chest abnormalities. In this work, we describe a dataset of  more than 100,000  chest X-ray scans that were retrospectively collected from two major hospitals in Vietnam. Out of this raw data, we release 18,000 images that were manually annotated by a total of 17 experienced radiologists with 22 local labels of rectangles surrounding abnormalities and 6 global labels of suspected diseases. The released dataset is divided into a training set of 15,000 and a test set of 3,000. Each scan in the training set was independently labeled by 3 radiologists, while each scan in the test set was labeled by the consensus of 5 radiologists. We designed and built a labeling platform for DICOM images to facilitate these annotation procedures. All images are made publicly available in DICOM format along with the labels of both the training set and the test set.  
\end{abstract}
\begin{document}

\flushbottom
\maketitle

\thispagestyle{empty}

\section*{Background \& Summary} \label{sec:introduction}
Computer-aided diagnosis (CAD) systems for chest radiographs (also referred to as Chest X-ray or CXR) have recently achieved great success thanks to the availability of large labeled datasets and the recent advances of high-performance supervised learning algorithms~\cite{rajpurkar2017chexnet,rajpurkar2018deep,irvin2019chexpert,majkowska2020chest,rajpurkar2020chexpedition,tang2020automated,pham2020interpreting}. 
Leveraging deep convolutional neural networks (CNN)~\cite{LeCunBH:2015}, these systems can reach the expert-level performance in classifying common lung diseases and related findings. Training a CNN heavily relies on high quality datasets of annotated images. However, it is costly and time-consuming to build such datasets due to several constraints: (1) medical data are hard to retrieve from hospitals or medical centers; (2) manual annotation by physicians is expensive; (3) the annotation of medical images requires a consensus of several expert readers to overcome human bias~\cite{razzak2018deep}; and (4) it lacks an efficient labeling framework to manage and annotate large-scale medical datasets.

Notable public datasets of CXR include ChestX-ray8, ChestX-ray14~\cite{wang2017chestx}, Padchest~\cite{bustos2019padchest}, CheXpert~\cite{irvin2019chexpert}, and MIMIC-CXR~\cite{johnson2019mimic}. ChestX-ray14, an extended version of ChestX-ray8, was released by the US National Institutes of Health (NIH), containing over 112,000 CXR scans from more than 30,000 patients. Without being manually annotated, this dataset poses significant issues related to the quality of its labels~\cite{rayner}. Padchest consists of more than 160,000 CXR images, 27\% of which were hand-labeled by radiologists with 174 different findings and 19 diagnoses. The rest of the dataset were labeled using a Natural Language Processing (NLP) tool. Recently released CheXpert provides more than 200,000 CXRs of 65,240 patients, which were labeled for the presence of 14 observations using an automated rule-based labeler that extracts keywords from medical reports. Adopting the same labeling mechanism, MIMIC-CXR contains 377,110 images in DICOM format along with free-text radiology reports. Table~\ref{existing-data} provides a summary of the aforementioned datasets together with other ones of moderate sizes, including JSRT~\cite{shiraishi2000development},
Indiana~\cite{demner2016preparing}, MC~\cite{jaeger2014two}, and SH~\cite{jaeger2014two}.
\begin{table}[ht]
\textsf{\scriptsize{
\caption{\textsf{An overview of existing public datasets for CXR interpretation.}}
\label{existing-data}
\setlength{\tabcolsep}{3pt}
\begin{tabular}{ p{90pt}|p{70pt}|p{70pt}|p{70pt}|p{70pt}|p{90pt}}
\hline
\hline
\cellcolor{Gray} Dataset &\cellcolor{Gray} \hspace*{0.5cm} Release year & \cellcolor{Gray}
\hspace*{0.5cm} \# findings & \cellcolor{Gray} \hspace*{0.5cm} \# samples & \cellcolor{Gray} \hspace*{0.2cm} Image-level labels & \cellcolor{Gray} \hspace*{0.85cm} Local labels\\
\hline
JSRT~\cite{shiraishi2000development} & \hspace*{1cm} 2000 & \hspace*{1cm} 1 & \hspace*{0.7cm} 247$^{(\triangleleft,\star)}$  & \hspace*{0.7cm} Available & \hspace*{1cm} Available\\
 \cellcolor{LightCyan} MC~\cite{jaeger2014two} & \cellcolor{LightCyan} \hspace*{1cm} 2014 & \cellcolor{LightCyan} \hspace*{1cm} 1 & \cellcolor{LightCyan} \hspace*{0.7cm} 138$^{(\triangleleft,\star)}$  & \cellcolor{LightCyan} \hspace*{0.7cm} Available & \cellcolor{LightCyan} \hspace*{1cm} N/A\\
    SH~\cite{jaeger2014two} &  \hspace*{1cm} 2014 & \hspace*{1cm}  1 &  \hspace*{0.7cm} 662$^{(\triangleleft,\star)}$  &  \hspace*{0.7cm} Available &  \hspace*{1cm} N/A\\
  \cellcolor{LightCyan} Indiana~\cite{demner2016preparing} & \cellcolor{LightCyan} \hspace*{1cm} 2016 &  \cellcolor{LightCyan} \hspace*{1cm} 10 & \cellcolor{LightCyan} \hspace*{0.7cm} 8,121$^{(\triangleleft,\star)}$ &  \cellcolor{LightCyan} \hspace*{0.7cm} Available & \cellcolor{LightCyan}  \hspace*{1cm} N/A\\
 ChestX-ray8~\cite{wang2017chestx} & \hspace*{1cm} 2017 &\hspace*{1cm}  8 &  \hspace*{0.7cm} 108,948$^{(\bullet)}$ & \hspace*{0.7cm} Available &  \hspace*{1cm} Available$^{(\dag)}$\\
 \cellcolor{LightCyan} ChestX-ray14~\cite{wang2017chestx}   & \cellcolor{LightCyan} \hspace*{1cm} 2017 &  \cellcolor{LightCyan} \hspace*{1cm} 14 & \cellcolor{LightCyan} \hspace*{0.7cm} 112,120$^{(\bullet)}$  &  \cellcolor{LightCyan} \hspace*{0.7cm} Available & \cellcolor{LightCyan} \hspace*{1cm} N/A \\
CheXpert~\cite{irvin2019chexpert} & \hspace*{1cm} 2019 & \hspace*{1cm} 14 & \hspace*{0.7cm} 224,316$^{(\bullet)}$ & \hspace*{0.7cm} Available & \hspace*{1cm} N/A\\
\cellcolor{LightCyan} Padchest~\cite{bustos2019padchest} & \cellcolor{LightCyan} \hspace*{1cm} 2019 &  \cellcolor{LightCyan} \hspace*{1cm} 193 & \cellcolor{LightCyan} \hspace*{0.7cm} 160,868$^{(\bullet}$ $^{,\star)}$  &   \cellcolor{LightCyan} \hspace*{0.7cm} Available & \cellcolor{LightCyan} \hspace*{1cm} N/A$^{(\dag\dag)}$\\
MIMIC-CXR~\cite{johnson2019mimic} & \hspace*{1cm} 2019 & \hspace*{1cm} 14 & \hspace*{0.7cm} 377,110$^{(\bullet)}$  & \hspace*{0.7cm} Available & \hspace*{1cm} N/A \\
\cellcolor{LightCyan} VinDr-CXR (ours)   & \cellcolor{LightCyan} \hspace*{1cm} 2020 & \cellcolor{LightCyan} \hspace*{1cm} 28 & \cellcolor{LightCyan} \hspace*{0.7cm} 18,000$^{(\star)}$  & \cellcolor{LightCyan} \hspace*{0.7cm} Available & \cellcolor{LightCyan} \hspace*{1cm} Available \\
\hline
\hline
\end{tabular}}}
\begin{flushleft}\scriptsize{$^{(\bullet)}$ Labeled by an NLP algorithm. $^{(\star)}$ Labeled by radiologists. $^{(\triangleleft)}$ Moderate-size datasets that are not applicable for training deep learning models. $^{(\dag)}$ A portion of the dataset (983 images) is provided with hand-labeled bounding boxes. $^{(\dag\dag)}$ 27\% of the dataset was manually annotated with encoded anatomical regions of the findings.} \end{flushleft}
\end{table}

Most existing CXR datasets depend on automated rule-based labelers that either use keyword matching (e.g. CheXpert~\cite{irvin2019chexpert} and NIH labelers~\cite{wang2017chestx}) or an NLP model (e.g. CheXbert~\cite{smit2020chexbert}) to extract disease labels from free-text radiology reports. These tools can produce labels on a large scale but, at the same time, introduce a high rate of inconsistency, uncertainty, and errors~\cite{rayner,oakden2020exploring}. These noisy labels may lead to the deviation of deep learning-based algorithms from reported performances when evaluated in a real-world setting~\cite{nagendran2020artificial}. Furthermore, the report-based approaches only associate a CXR image with one or several labels in a predefined list of findings and diagnoses without identifying their locations. There are a few CXR datasets that include annotated locations of abnormalities but they are either too small for training deep learning models (JSRT) or not detailed enough (PadChest). The interpretation of a CXR is not all about image-level classification; it is even more important, from the perspective of a radiologist, to localize the abnormalities on the image. This partly explains why the applications of CAD systems for CXR in clinical practice are still very limited.    

In an effort to provide a large CXR dataset with high-quality labels for the research community, we have built the VinDr-CXR dataset from more than 100,000 raw images in DICOM format that were retrospectively collected from the Hospital 108 (H108) and the Hanoi Medical University Hospital (HMUH), two of the largest hospitals in Vietnam. The published dataset consists of 18,000 postero-anterior (PA) view CXR scans that come with both the localization of critical findings and the classification of common thoracic diseases. These images were annotated by a group of 17 radiologists with at least 8 years of experience for the presence of 22 critical findings (local labels) and 6 diagnoses (global labels); each finding is localized with a bounding box. The local and global labels correspond to the ``Findings'' and ``Impressions'' sections, respectively, of a standard radiology report.  We divide the dataset into two parts: the training set of 15,000 scans and the test set of 3,000 scans. Each image in the training set was independently labeled by 3 radiologists, while the annotation of each image in the test set was even more carefully treated and obtained from the consensus of 5 radiologists. The labeling process was performed via an in-house system called VinDr Lab~\cite{VinDr-Lab}, which was built on top of a Picture Archiving and Communication System (PACS). All DICOM images and the labels of both the training set and the test set are released. A slightly modified version of this dataset was used to organize the VinBigData Chest Xray Abnormalities Detection Challenge on the Kaggle platform (\href{https://www.kaggle.com/c/vinbigdata-chest-xray-abnormalities-detection/}{https://www.kaggle.com/c/vinbigdata-chest-xray-abnormalities-detection/}).

VinDr-CXR, to the best of our knowledge, is currently the largest public CXR dataset with radiologist-generated annotations in both training and test sets. We believe the dataset will accelerate the development and evaluation of new machine learning models for both localization and classification of thoracic lesions and diseases on CXR scans. 

 \section*{Methods}
\begin{figure}[H]
\centerline{\includegraphics[width=\linewidth]{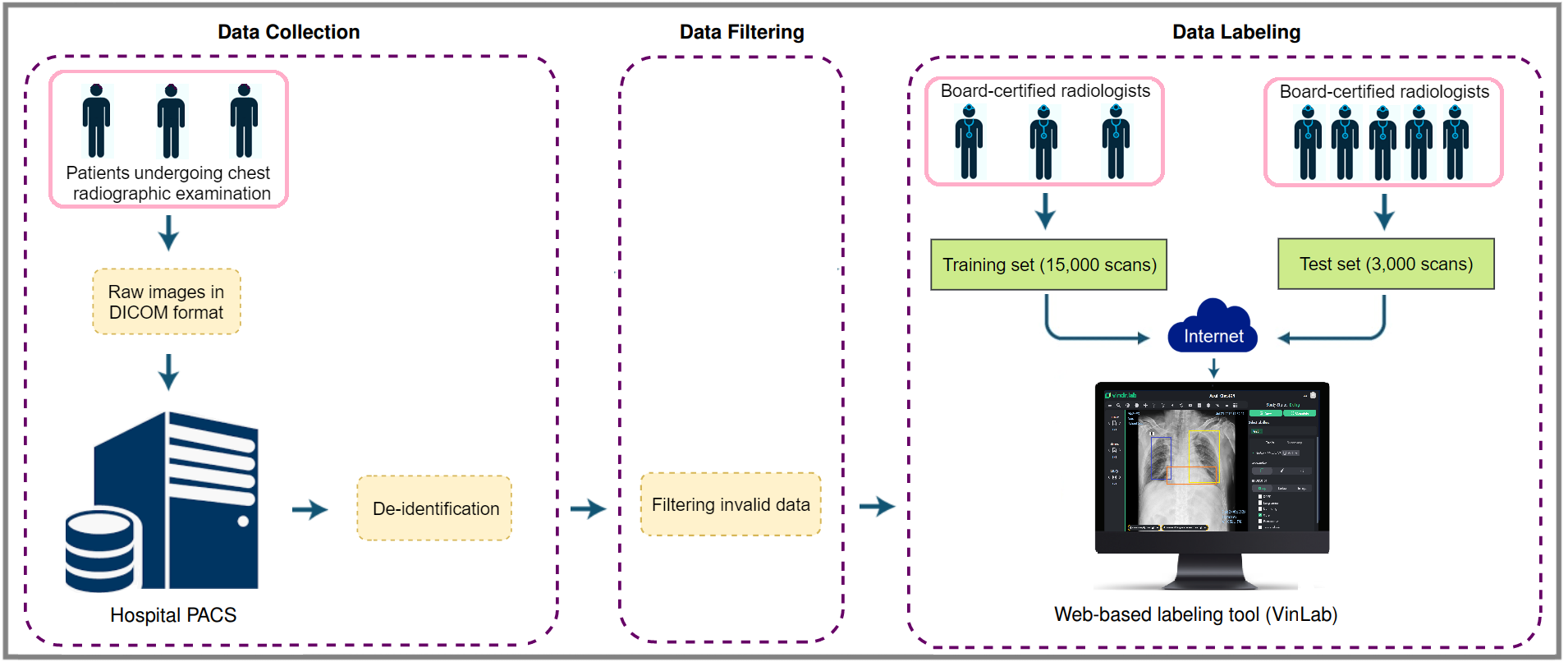}}
\caption{{\textsf{\small{The flow of creating VinDr-CXR dataset: (1) raw images in DICOM format were collected retrospectively from the hospital's PACS and got de-identified to protect patient's privacy; (2) invalid files, such as images of other modalities, other body parts, low quality, or incorrect orientation, were automatically filtered out by a CNN-based classifier; (3) A web-based labeling tool, VinDr Lab, was developed to store, manage, and remotely annotate DICOM data: each image in the training set of 15,000 images was independently labeled by a group of 3 radiologists and each image in the test set of 3,000 images was labeled by the consensus of 5 radiologists.}}}}
\label{fig:data_colection}
\end{figure}

The building of VinDr-CXR dataset, as visualized in Figure~\ref{fig:data_colection}, is divided into three main steps: (1) data collection, (2) data filtering, and (3) data labeling. Between 2018 and 2020, we retrospectively collected more than 100,000 CXRs in DICOM format from local PACS  servers of two  hospitals in Vietnam, HMUH and  H108. Imaging data were acquired from a wide diversity of scanners from well-known medical equipment manufacturers, including Phillips, GE, Fujifilm, Siemens, Toshiba, Canon, Samsung, and Carestream. The ethical clearance of this study was approved by the Institutional Review Boards (IRBs) of the HMUH and H108 before the study started. The need for obtaining informed patient consent was waived because this retrospective study did not impact clinical care or workflow at these two hospitals and all patient-identifiable information in the data has been removed.

\subsection*{Data de-identification}
\label{sec:dataset}
To protect patient's privacy~\cite{article_8}, all personally identifiable information associated with the images has been removed or replaced with random values. Specifically, we ran a Python script that removes all DICOM tags of protected health information (PHI)~\cite{isola2019protected} such as: patient's name, patient's date of birth, patient ID, or acquisition time and date, etc. We only retained a limited number of DICOM attributes that are necessary for processing raw images. The entire list of retained attributes is shown in Table~1 (\textcolor{blue}{supplementary materials}). Next, a simple algorithm was implemented to automatically remove textual information appearing on the image data (i.e. pixel annotations that could include patient's identifiable information). The resulting images were then manually verified  to make sure all texts were removed before they were digitally sent out of the hospitals' systems.

\subsection*{Data filtering}
\label{sec:dataset}
\begin{figure}[!t]
\centerline{\includegraphics[width=17.8cm,height=7cm]{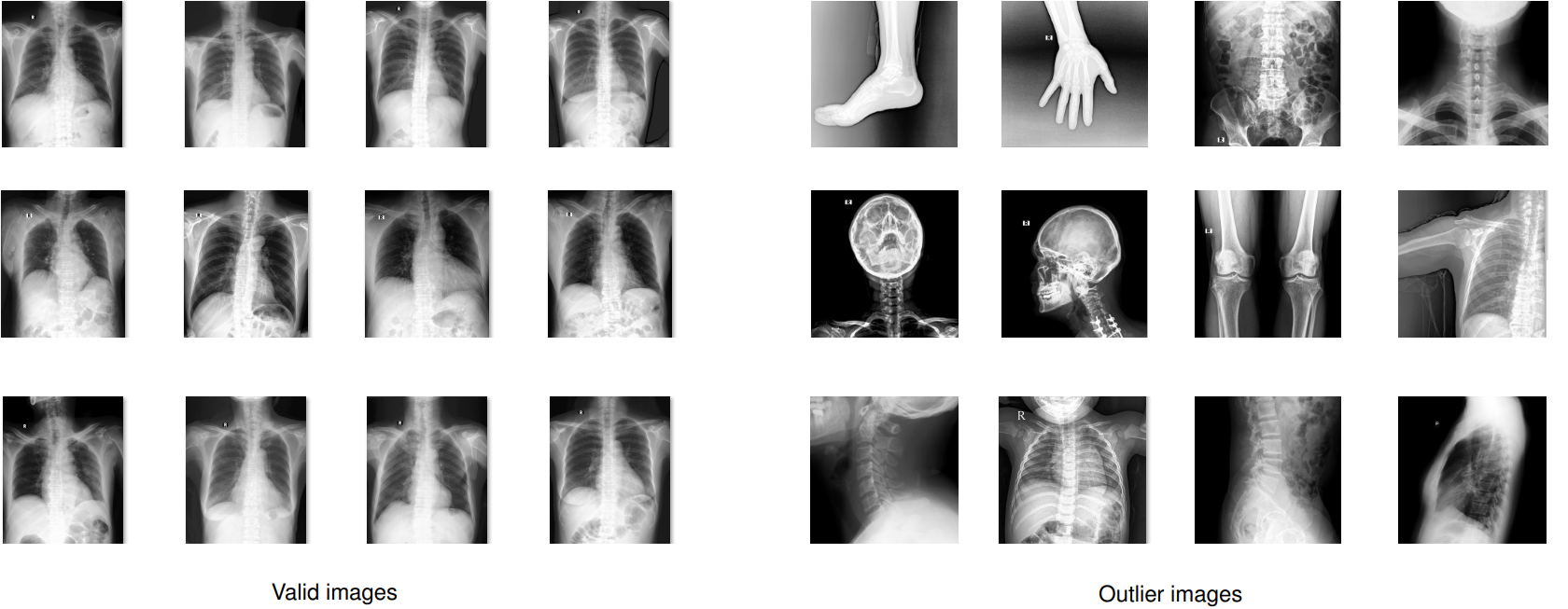}}
\caption{{\textsf{\small{Examples of valid (\textbf{left}) and invalid (\textbf{right}) CXR scans. A CNN-based classifier was trained and used to automatically filter outliers; only valid PA-view CXRs of adults were retained for labeling.}}}}
\label{outliers}
\end{figure}

The collected raw data was mostly of \emph{adult PA-view CXRs}, but also included a significant amount of outliers such as images of body parts other than chest (due to mismatched DICOM tags), pediatric scans,  low-quality images, or lateral CXRs. Examples of these images are shown in Figure~\ref{outliers}. All outliers were automatically excluded from the dataset using a binary classifier, which is a light-weight convolutional neural network (CNN). The training procedure of this classifier is out of the scope of this paper. 

\subsection*{Data labeling}
\begin{figure}[ht]
\centerline{\includegraphics[width=\linewidth]{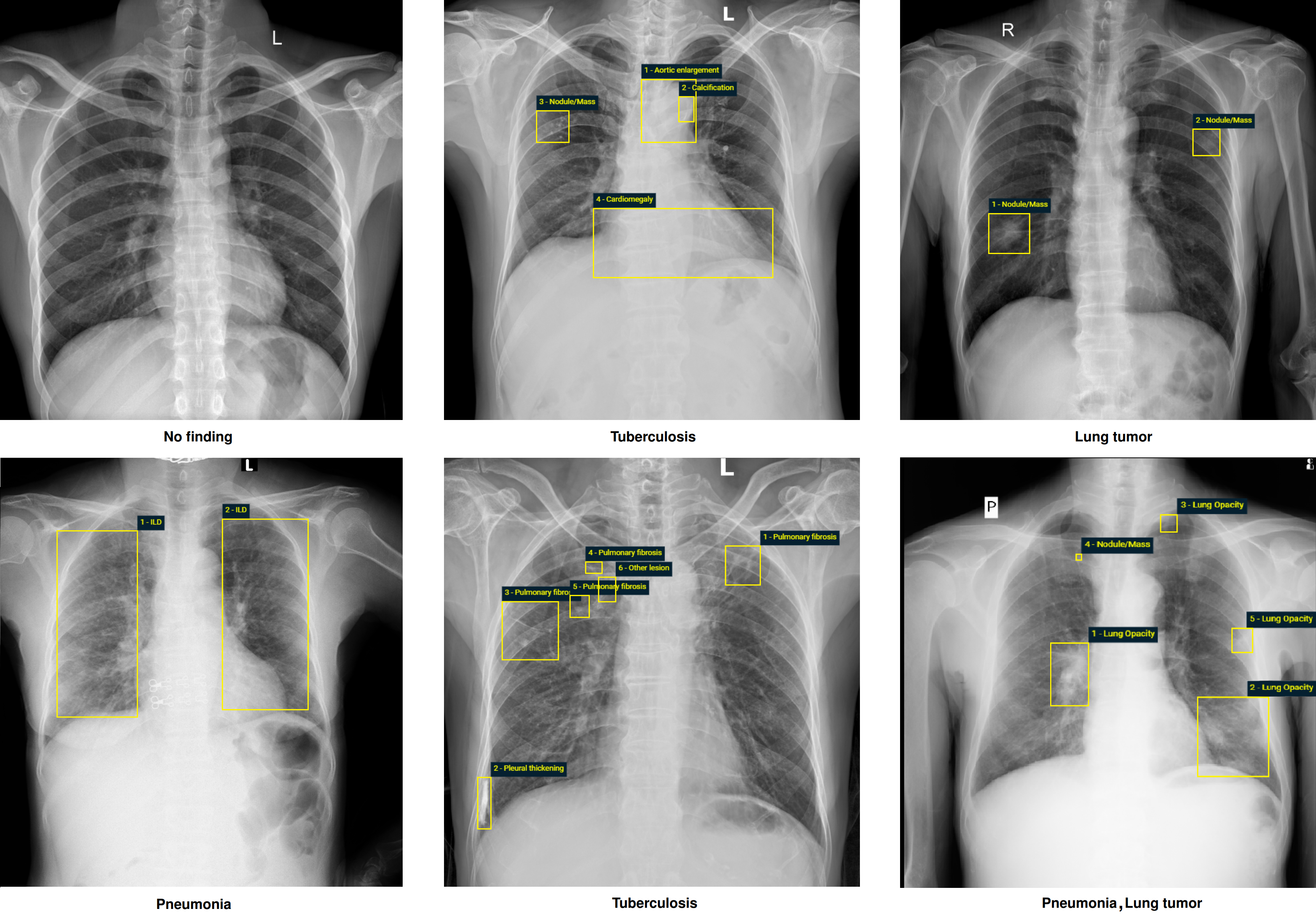}}
\caption{{\textsf{\small{Examples of CXRs with radiologist's annotations. Abnormal findings (local labels) marked by radiologists are plotted on the original images for visualization purpose. The global labels are in bold and listed at the bottom of each example. Better viewed on a computer and zoomed in for details.}}}}
\label{representive_cases}
\end{figure}

The VinDr-CXR dataset was labeled for a total of 28 findings and diagnoses in adult cases: (1) Aortic enlargement, (2) Atelectasis, (3) Cardiomegaly, (4) Calcification, (5) Clavicle fracture, (6) Consolidation, (7) Edema, (8) Emphysema, (9) Enlarged PA, (10) Interstitial lung disease (ILD), (11) Infiltration, (12) Lung cavity, (13) Lung cyst, (14) Lung opacity, (15) Mediastinal shift, (16) Nodule/Mass, (17) Pulmonary fibrosis, (18) Pneumothorax, (19) Pleural thickening, (20) Pleural effusion, (21) Rib fracture, (22) Other lesion, (23) Lung  tumor, (24) Pneumonia, (25) Tuberculosis, (26) Other diseases, (27) Chronic obstructive pulmonary disease (COPD), and (28) No finding. These labels were divided into 2 categories: local labels (1-22) and global labels (23-28). The local labels should be marked with bounding boxes that localize the findings, while the global labels should reflect the diagnostic impression of the radiologist. The definition of each label is detailed in Table~2 (\textcolor{blue}{supplementary materials}). This list of labels was suggested by a committee of the most experienced radiologists from the two hospitals. The selection of these labels took into account two factors: first, they are prevalent and second, they can be differentiated on CXRs.  Figure~\ref{representive_cases} illustrates several samples with both local and global labels annotated by radiologists.

To facilitate the labeling process, we designed and built a web-based framework called VinDr Lab and had a team of 17 experienced radiologists remotely annotate the data. All the radiologists participating in the labeling process were certified in diagnostic radiology and received healthcare profession certificates from the Vietnamese Ministry of Health. A set of 18,000 CXRs were randomly chosen from the filtered data, of which 15,000 scans (normal: 10,606 studies, abnormal:  4394 studies) serve as the training set and the remaining 3,000 (normal: 2052 studies, abnormal: 948 studies)form the test set. Each sample in the training set was assigned to 3 radiologists for annotating in a blind fashion. Additionally, all of the participating radiologists were blinded to relevant clinical information. For the test set, 5 radiologists were involved in a two-stage labeling process. During the first stage, each image was independently annotated by 3 radiologists. In the second stage, 2 other radiologists, who have a higher level of experience, reviewed the annotations of the 3 previous annotators and communicated with each other in order to decide the final labels. The disagreements among initial annotators were carefully discussed and resolved by the 2 reviewers. Finally, the consensus of their opinions will serve as reference ground-truth. 

\begin{figure}[H]
\centerline{\includegraphics[width=14.5cm,height=9cm]{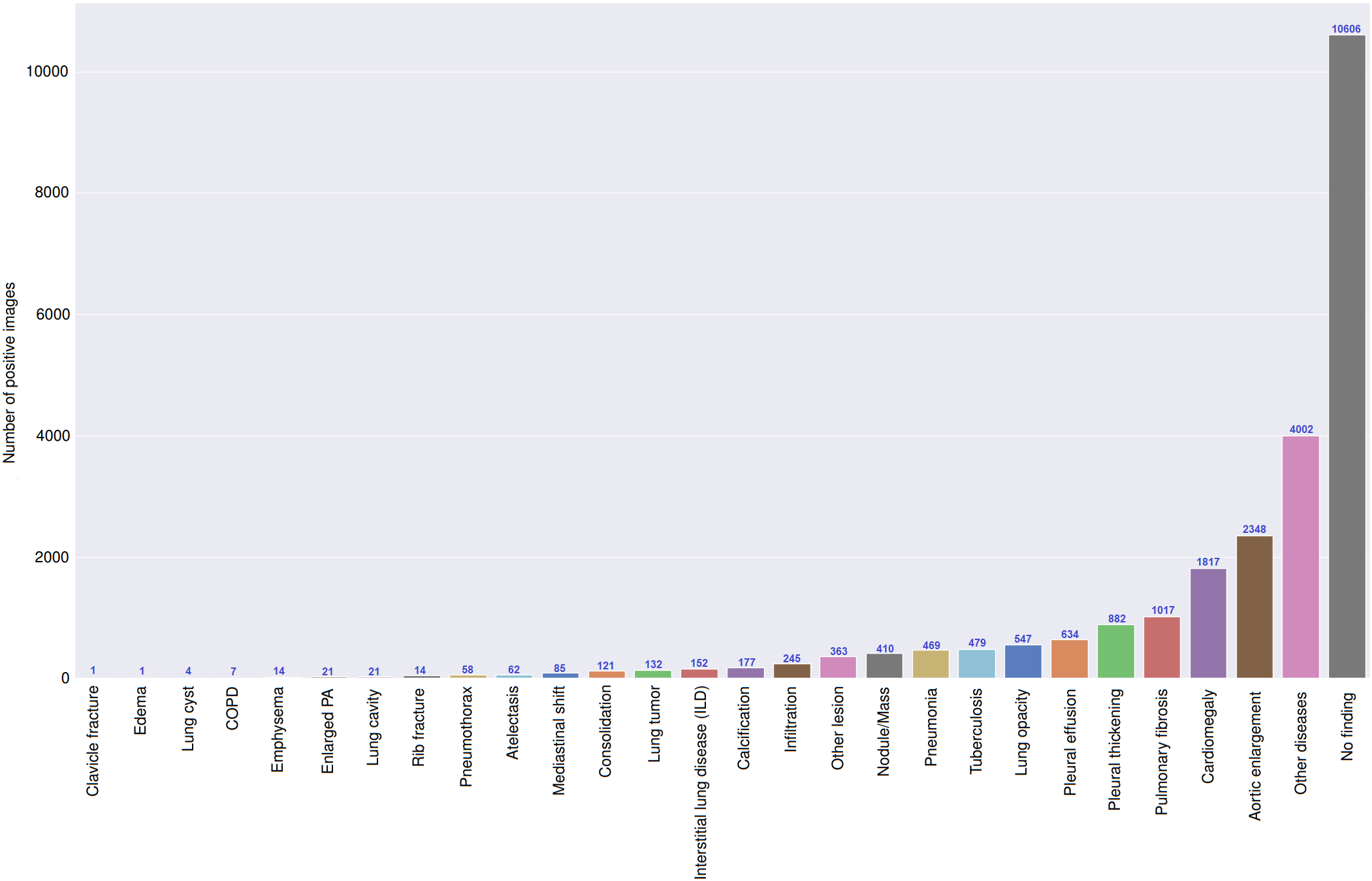}}
\caption{{\textsf{\small{Distribution of findings and pathologies on the training set of VinDr-CXR.}}}}
\label{label-distribution}
\end{figure}

Once the labeling was completed, the labels of 18,000 CXRs were exported in JavaScript Object Notation (JSON) format. We then parsed their contents and organized the annotations in the form of a single comma-separated values (CSV) file. As a result, we provided a single CSV file that contains labels, bounding box coordinates, and their corresponding image IDs. For the training set, each sample comes with the annotations of three different radiologists. For the test set, we only provide the consensus labels of the five radiologists. The data characteristics, including patient demographic and the prevalence of each finding or pathology, are summarized in Table~\ref{data-characs}. The distribution of all labels in the training set is drawn in Figure~\ref{label-distribution}. We have released all images together with the labels of the training set and the test set.

\begin{table}[ht]
\centering
\textsf{\scriptsize{
\caption{\textsf{Dataset characteristics}}
\label{data-characs}
\setlength{\tabcolsep}{3pt}
\begin{tabular}{p{15pt} | p{180pt}|p{100pt}|p{120pt}}
\hline
\hline
& \cellcolor{Gray} Characteristics& \cellcolor{Gray}
Training set & \cellcolor{Gray}
Test set  \\
\hline
  \parbox[t]{2mm}{\multirow{10}{*}{\rotatebox[origin=c]{90}{\textbf{Collection statistics}}}} & &  \\
  &  \cellcolor{LightCyan} Years & \cellcolor{LightCyan} 2018 to 2020 & \cellcolor{LightCyan} 2018 to 2020\\
&  Number of scans & 15,000 & 3,000 \\
&  \cellcolor{LightCyan} Number of human annotators per scan  & \cellcolor{LightCyan} 3 & \cellcolor{LightCyan} 5  \\
&  Image size (pixel$\times$pixel, median) &  2788 $\times$ 2446 &  2748 $\times$ 2394 \\
&  \cellcolor{LightCyan} Age (years, median)*  & \cellcolor{LightCyan} 43.77 & \cellcolor{LightCyan} 31.80  \\
&   Male (\%)* &  52.21  & 55.90    \\
&  \cellcolor{LightCyan} Female (\%)* & \cellcolor{LightCyan} 47.79  & \cellcolor{LightCyan} 44.10  \\
&  Data size (GB) &   161 &   31.3 \\
& & &  \\
\hline
\parbox[t]{3mm}{\multirow{22.5}{*}{\rotatebox[origin=c]{90}{\textbf{Local labels}}}}  & &   \\
& \cellcolor{LightCyan} 1. Aortic enlargement (\%)  & \cellcolor{LightCyan} 2348 (15.65\%)  &  \cellcolor{LightCyan}  220 (7.33\%) \\
&                       2. Atelectasis (\%)         &  62 (0.41\%)                          & 86 (2.87\%)  \\
& \cellcolor{LightCyan} 3. Cardiomegaly (\%)        & \cellcolor{LightCyan} 1817 (12.11\%)  & \cellcolor{LightCyan}  309 (10.30\%) \\
&                       4. Calcification (\%)       & 177 (1.18\%)                          & 194 (6.47\%) \\
& \cellcolor{LightCyan} 5. Clavicle fracture (\%)   & \cellcolor{LightCyan} 1 (0.01\%)      & \cellcolor{LightCyan} 2 (0.07\%) \\
&                       6. Consolidation (\%)       &  121 (0.81\%)                         & 96 (3.20\%) \\
& \cellcolor{LightCyan} 7. Edema (\%)               & \cellcolor{LightCyan} 1 (0.01\%)      & \cellcolor{LightCyan}  0 (0\%) \\
&                       8. Emphysema (\%)           &  14 (0.09\%)                          & 3 (0.1\%) \\
& \cellcolor{LightCyan} 9. Enlarged PA (\%)         & \cellcolor{LightCyan} 21 (0.14\%)     & \cellcolor{LightCyan} 8 (0.27\%) \\
&                       10. Interstitial lung disease (ILD) (\%)&  152 (1.01\%)             & 221 (7.37\%) \\
& \cellcolor{LightCyan} 11. Infiltration (\%)       & \cellcolor{LightCyan} 245 (1.63\%)    & \cellcolor{LightCyan} 58 (1.93\%) \\
&                       12. Lung cavity (\%)        &  21 (0.14\%)                          & 9 (0.30\%) \\
& \cellcolor{LightCyan} 13. Lung cyst (\%)          & \cellcolor{LightCyan}  4 (0.03\%)     & \cellcolor{LightCyan} 2 (0.07\%) \\
&                       14. Lung opacity (\%)       &   547 (3.65\%)                        & 84 (2.80\%) \\
& \cellcolor{LightCyan} 15. Mediastinal shift (\%)  & \cellcolor{LightCyan} 85 (0.57\%)     & \cellcolor{LightCyan} 20 (0.67\%) \\
&                       16. Nodule/Mass (\%)        &  410 (2.73\%)                         & 176 (5.87\%) \\
& \cellcolor{LightCyan} 17. Pulmonary fibrosis (\%) & \cellcolor{LightCyan} 1017 (6.78\%)   & \cellcolor{LightCyan} 217 (7.23\%) \\
&                       18. Pneumothorax (\%)       &   58 (0.39\%)                         & 18 (0.60\%) \\
& \cellcolor{LightCyan} 19. Pleural thickening (\%) &  \cellcolor{LightCyan} 882 (5.88\%)   & \cellcolor{LightCyan} 169 (5.63\%) \\
&                       20. Pleural effusion (\%)   &  634 (4.23\%)                         & 111 (3.70\%) \\
& \cellcolor{LightCyan} 21. Rib fracture (\%)       & \cellcolor{LightCyan} 41 (0.27\%)     & \cellcolor{LightCyan} 11 (0.37\%) \\
&                       22. Other lesion (\%)       &  363 (2.42\%)                         & 94 (3.13\%) \\
& & & 
\\
\hline 
\parbox[t]{2mm}{\multirow{8}{*}{\rotatebox[origin=c]{90}{\textbf{Global labels}}}} & &  \\
& \cellcolor{LightCyan} 23. Lung  tumor (\%)        & \cellcolor{LightCyan}  132 (0.88\%)   & \cellcolor{LightCyan} 80 (2.67\%) \\
& 24. Pneumonia (\%)                                &  469 (3.13\%)                         &  246 (8.20\%) \\
& \cellcolor{LightCyan} 25. Tuberculosis (\%)       & \cellcolor{LightCyan} 479 (3.19\%)    & \cellcolor{LightCyan} 164 (5.47\%)  \\
&  26. Other diseases (\%)                          &   4002 (26.68\%)                      &  657 (21.90\%) \\
& \cellcolor{LightCyan} 27. COPD (\%)               & \cellcolor{LightCyan} 7 (0.05\%)      & \cellcolor{LightCyan}  2 (0.07\%) \\
&  28. No finding (\%)                              & 10606 (70.71\%)                       &  2051 (68.37\%)  \\
\hline
\hline
\end{tabular}
}}
\begin{flushleft} \scriptsize{ Note: the numbers of positive labels were reported based on the majority vote of the participating radiologists. (*) The calculations were only based on the CXR scans where patient's sex and age were known.} \end{flushleft}
\end{table}

\section*{Data Records}
\label{sec:data_records}
The VinDr-CXR dataset has been submitted to PhysioNet (\href{https://physionet.org/content/vindr-cxr/1.0.0/}{https://physionet.org/content/vindr-cxr/1.0.0/}) for public download. We provide all imaging data and the corresponding ground truth labels for both the training and test sets. The images were organized into two folders, one for training and the other one for testing. Each image has a unique, anonymous identifier which was encoded from the value of the SOP Instance UID provided by the DICOM tag (0008,0018). The encoding process was supported by the Python \verb|hashlib| module (see \hyperref[sec:code]{Code Availability}). The radiologists' local annotations of the training set were provided in a CSV file, \verb|annotations_train.csv|. Each row of the table represents a bounding box with the following attributes: image ID (\verb|image_id|), radiologist ID (\verb|rad_id|), label's name (\verb|class_name|), and bounding box coordinates (\verb|x_min|, \verb|y_min|, \verb|x_max|, \verb|y_max|). Here, \verb|rad_id| encodes the identities of the 17 radiologists, (\verb|x_min|, \verb|y_min|) are the coordinates of the box's upper left corner, and (\verb|x_max|, \verb|y_max|) are the coordinates of the lower-right corner. Meanwhile, the image-level labels of the training set were stored in different CSV file, \verb|image_labels_train.csv|, with the following fields: Image ID (\verb|image_id|), radiologist ID (\verb|rad_ID|), and labels (\verb|labels|) for both the findings and diagnoses. Specifically, each image ID goes with vector of multiple labels corresponding to different pathologies, in which positive ones were encoded with ``1'' and negative ones were encoded with ``0''. Similarly, the bounding-box annotations and the image-level labels of the test set were recorded in \verb|annotations_test.csv| and \verb|image_labels_test.csv|, respectively. The only difference is that each row in the CSV files of the test set was not associated with a radiologist ID.   

\section*{Technical Validation}

The data de-identification was controlled. In particular, all DICOM meta-data was parsed and manually reviewed to ensure that all individually identifiable health information of the patients has been removed to meet the U.S. HIPAA (\href{https://www.hhs.gov/hipaa/for-professionals/privacy/laws-regulations/index.html}{https://www.hhs.gov/hipaa/for-professionals/privacy/laws-regulations/index.html}), the European GDPR~(\href{https://gdpr-info.eu/}{https://gdpr-info.eu/}), as well as the local privacy laws. Pixel values of all CXR scans were also carefully examined. All images were  manually reviewed case-by-case by a team of 10 human readers. During this review process, a small number of images containing private textual information that had not been removed by our algorithm was excluded from the dataset. The manual review process also helped identify and discard outlier samples that the CNN-based classifier was not able to detect. To control the quality of the labeling process, we developed a set of rules underlying VinDr Lab for automatic verification of radiologist-generated labels. These rules prevent annotators from mechanical mistakes like forgetting to choose global labels or marking lesions on the image while choosing ``No finding'' as the global label. To ensure the complete blindness among annotators, the images were randomly shuffled before being assigned to each of them.

\section*{Usage Notes}
To download the dataset, users are required to accept a Date Usage Agreement (DUA) called PhysioNet Credentialed Health Data License 1.5.0 (\href{https://physionet.org/content/vindr-cxr/view-license/1.0.0/)}{https://physionet.org/content/vindr-cxr/view-license/1.0.0/}). By accepting the DUA, users agree that they will not share the data and that the dataset can be used for scientific research and educational purposes only and will not attempt to re-identify any patients, institutions or hospitals. For any publication that explores this resource, the authors must cite this original paper. We also encourage such  authors to release their code and models, which will help the community to reproduce experiments and to boost the research in the field of medical imaging.

\section*{Code Availability}
\label{sec:code}
The code used for loading and processing DICOM images is based on the following open-source repositories: Python 3.7.0 (\href{https://www.python.org/}{https://www.python.org/}); Pydicom 1.2.0 (\href{https://pydicom.github.io/}{https://pydicom.github.io/}); OpenCV-Python 4.2.0.34 (\href{https://pypi.org/project/opencv-python/}{https://pypi.org/project/opencv-python/}); and Python hashlib (\href{https://docs.python.org/3/library/hashlib.html}{https://docs.python.org/3/library/hashlib.html}). The code for data de-identification and outlier detection was made publicly available at \href{https://github.com/vinbigdata-medical/vindr-cxr}{https://github.com/vinbigdata-medical/vindr-cxr}.

\bibliography{ref}

\section*{Acknowledgements}

The authors would like to acknowledge the Hanoi Medical University Hospital and the  Hospital 108 for providing us access to their image databases and for agreeing to make the VinDr-CXR dataset publicly available. We are especially thankful to all of our collaborators, including  radiologists, physicians, and technicians, who participated in the data collection and labeling process. 


\section*{Author contributions}

H.Q.N., K.L., and L.L. designed the study; H.Q.N., Nghia T. Nguyen, M.D., and V.V. designed the labeling framework; H.H.P. and D.B.N. performed the data de-identification; H.H.P. developed the algorithm for outlier filtering; D.T., D.B.N., D.T.N., and Nhan T. Nguyen conducted the data acquisition and analysis; K.L, L.L, D.L., C.P., H.T., D.D., C.D., L.D., C.N., B.N, Q.N., A.H., H.N.P., A.N., and P.H. annotated data and made comments to improve the labeling tools; H.Q.N., and H.H.P. wrote the paper; all authors reviewed the manuscript.

\section*{Competing interests} 

This work was funded by the Vingroup JSC. The funder had no role in study design, data collection and analysis, decision to publish, or preparation of the manuscript.

\newpage 


\section*{Supplementary materials \label{supp_materi}}

\begin{table*}[hbt!]
\textsf{\scriptsize{
\caption{\textsf{Definition of findings and diseases used in the study.}}
\label{label-defination}
\centering
\setlength{\tabcolsep}{3pt}
\begin{tabular}{p{10pt} | p{110pt}|p{300pt}}
\\
\hline
& \textbf{Pathology label} & 
\hspace{5cm}\textbf{Definition}\\
\hline
\parbox[t]{3mm}{\multirow{60}{*}{\rotatebox[origin=c]{90}{\textbf{Local Label}}}}  &   \\
& 1. Aortic enlargement & An abnormal bulge that occurs in the wall of the major blood vessel. \\
& \\
& 2. Atelectasis &  Collapse of a part of the lung due to a decrease in the amount of air in the alveoli resulting in volume loss and increased density.\\
& \\
& 3. Cardiomegaly &   Enlargement of the heart, occurs when the heart of an adult patient is larger than normal and the cardiothoracic ratio is greater than 0.5.\\
& \\
& 4. Calcification & Deposition of calcium salts in the lung. \\
& \\
& 5. Clavicle fracture & A break in the collarbone.\\
& \\
& 6. Consolidation &  Any pathologic process that fills the alveoli with fluid, pus, blood, cells (including tumor cells) or other substances resulting in lobar, diffuse or multifocal ill-defined opacities. \\
& \\
& 7. Edema & Fluid accumulation in the tissur and air space of the lungs.\\
& \\
& 8. Emphysema  & A condition of the lung characterized by an abnormal increase in the size of air spaces distal to the terminal bronchioles.\\
& \\
& 9. Enlarged PA & Dilatation of the pulmonary artery - a defect characterized by a wider than normal main pulmonary artery.\\
& \\
& 10. Interstitial lung disease (ILD) &  Involvement of the supporting tissue of the lung parenchyma resulting in fine or coarse reticular opacities or small nodules. \\
& \\
& 11. Infiltration &  An abnormal substance that accumulates gradually within cells or body tissues or any substance or type of cell that occurs within or spreads as through the interstices (interstitium and/or alveoli) of the lung, that is foreign to the lung, or that accumulates in greater than normal quantity within it. \\
& \\
& 12. Lung cavity & Thick-walled abnormal gas-filled spaces within the lung. They are usually associated with a nodule, mass, or area of consolidation. A fluid level within the space may be present.\\
& \\
& 13. Lung cyst & Lung cysts refer to round, thin-walled, low attenuation spaces/lucencies in the lung.\\
& \\
& 14. Lung opacity &  Any abnormal focal or generalized opacity or opacities in lung fields (blanket tag including but not limited to consolidation, cavity, fibrosis, nodule, mass, calcification, interstitial thickening, etc.). \\
& \\
& 15. Mediastinal shift & The deviation of the mediastinal structures towards one side of the chest cavity.\\
& \\
& 16. Nodule/Mass &  Any space occupying lesion either solitary or multiple.\\
& \\
& 17. Pulmonary fibrosis &  An excess of fibrotic tissue in the lung.\\
& \\
& 18. Pneumothorax &  The presence of gas (air) in the pleural space. \\
& \\
& 19. Pleural thickening &  Any form of thickening involving either the parietal or visceral pleura. \\
& \\
& 20. Pleural effusion & Abnormal accumulations of fluid within the pleural space. \\
& \\
& 21. Rib fracture & A common injury that occurs when one of the bones in your rib cage breaks or cracks.\\
& \\
& 22. Other lesion &  Other lesions that are not on the list of findings or abnormalities mentioned above. \\
& & 
\\
\hline 
\parbox[t]{2mm}{\multirow{10}{*}{\rotatebox[origin=c]{90}{\textbf{Global labels}}}} & &  \\
& 23. Lung tumor &  The result of abnormal rates of cell division or cell death in lung tissue, or in the airways that lead to the lungs. \\
& \\
& 24. Pneumonia &  An infection that inflames the air sacs in one or both lungs. \\
& \\
& 25. Tuberculosis & Any sign suggesting pulmonary or extrapulmonary tuberculosis.\\
& \\
& 26. Other diseases &  Other diseases that are not on the list of diseases mentioned above.\\
& \\
& 27. COPD  & Chronic obstructive pulmonary disease (COPD) is defined as a condition characterized by persistent airflow limitation that is usually progressive and associated with an enhanced chronic inflammatory response in the airways and the lung to noxious particles or gases.\\
& \\
& 28. No finding &  The absence of all pathologies from the chest radiograph. \\
& & 
\\
\hline
\end{tabular}
}}
\end{table*}

\begin{table*}
\centering
\textsf{\scriptsize{
\caption{\textsf{The list of DICOM tags that were retained for loading and processing raw images. All other tags were removed for protecting patient privacy. Details about all these tags can be found from DICOM Standard Browser at \href{https://dicom.innolitics.com/ciods}{https://dicom.innolitics.com/ciods}.}}
\label{dicom_tags_retained}
\setlength{\tabcolsep}{3pt}
\begin{tabular}{p{130pt}|p{150pt}|p{200pt}}
\\
\hline
\hspace*{1cm} \textbf{DICOM Tag} & 
\hspace*{1cm}\textbf{Attribute Name} & \textbf{Description} \\
\hline
& & \\
\hspace*{1cm} (0010, 0040) & \hspace*{1cm} Patient's Sex & Sex of the named patient.\\
& & \\
\hspace*{1cm} (0010, 1010) & \hspace*{1cm} Patient's Age & Age of the patient.\\
& & \\
\hspace*{1cm} (0010, 1020) & \hspace*{1cm} Patient's Size & Length or size of the patient, in meters.\\
& \\
\hspace*{1cm} (0010, 1030) & \hspace*{1cm} Patient's Weight & Weight of the patient, in kilograms.\\
& \\
\hspace*{1cm} (0028, 0010) & \hspace*{1cm} Rows & Number of rows in the image.\\
& \\
\hspace*{1cm} (0028, 0011) & \hspace*{1cm} Columns & Number of columns in the image.\\
& \\
\hspace*{1cm} (0028, 0030) & \hspace*{1cm} Pixel Spacing & Physical distance in the patient between the center of each pixel, specified by a numeric pair - adjacent row spacing (delimiter) adjacent column spacing in mm. \\
& \\
\hspace*{1cm} (0028, 0034) & \hspace*{1cm} Pixel Aspect Ratio & Ratio of the vertical size and horizontal size of the pixels in the image specified by a pair of integer values where the first value is the vertical pixel size, and the second value is the horizontal pixel size.\\
& \\
\hspace*{1cm} (0028, 0100) & \hspace*{1cm} Bits Allocated & Number of bits allocated for each pixel sample. Each sample shall have the same number of bits allocated. \\
& \\
\hspace*{1cm} (0028, 0101) & \hspace*{1cm} Bits Stored & Number of bits stored for each pixel sample. Each sample shall have the same number of bits stored.\\
& \\
\hspace*{1cm} (0028, 0102) & \hspace*{1cm} High Bit & Most significant bit for pixel sample data. Each sample shall have the same high bit. \\
& \\
\hspace*{1cm} (0028, 0103) & \hspace*{1cm} Pixel Representation & Data representation of the pixel samples. Each sample shall have the same pixel representation.\\
& \\
\hspace*{1cm} (0028, 0106) & \hspace*{1cm} Smallest Image Pixel Value & The minimum actual pixel value encountered in this image.\\
& \\
\hspace*{1cm} (0028, 0107) & \hspace*{1cm} Largest Image Pixel Value & The maximum actual pixel value encountered in this image.\\
& \\
\hspace*{1cm} (0028, 1050) & \hspace*{1cm} Window Center & Window center for display.\\
& \\ 
\hspace*{1cm} (0028, 1051) & \hspace*{1cm} Window Width & Window width for display.\\
& \\
\hspace*{1cm} (0028, 1052) & \hspace*{1cm} Rescale Intercept & The value b in relationship between stored values (SV) and the output units specified in Rescale Type (0028,1054). Each output unit is equal to m*SV + b.\\
& \\
\hspace*{1cm} (0028, 1053) & \hspace*{1cm} Rescale Slope & Value of m in the equation specified by Rescale Intercept (0028,1052).\\
& \\
\hspace*{1cm} (7FE0, 0010) & \hspace*{1cm} Pixel Data & A data stream of the pixel samples that comprise the image.\\
& \\
\hspace*{1cm} (0028, 0004) & \hspace*{1cm} Photometric Interpretation & Specifies the intended interpretation of the pixel data.\\
& \\
\hspace*{1cm} (0028, 2110) & \hspace*{1cm} Lossy Image Compression & Specifies whether an image has undergone lossy compression (at a point in its lifetime).\\
&\\
\hspace*{1cm} (0028, 2114) & \hspace*{1cm} Lossy Image Compression Method & A label for the lossy compression method(s) that have been applied to this image.\\
&\\
\hspace*{1cm} (0028, 2112) & \hspace*{1cm} Image Compression Ratio & Describes the approximate lossy compression ratio(s) that have been applied to this image.\\
&\\
\hspace*{1cm} (0028, 0002) & \hspace*{1cm} Samples per Pixel & Number of samples (planes) in this image. \\
&\\
\hspace*{1cm} (0028, 0008) & \hspace*{1cm} Number of Frames & Number of frames in a multi-frame image.\\
&\\
\hline
\end{tabular}
}}

\end{table*}

\end{document}